# ISRU Implications for Lunar and Martian Plume Effects

Philip T. Metzger[1]
*NASA, Kennedy Space Center, FL 32899*

Xiaoyi Li[2]
*Analex Corp., Kennedy Space Center, FL 32899*

and

Christopher D. Immer[3] and John E. Lane[4]
*ASRC Aerospace, Kennedy Space Center, FL 32899*

**Experiments, analyses, and simulations have shown that the engine exhaust plume of a Mars lander large enough for human spaceflight will create a deep crater in the martian soil, blowing ejecta to approximately 1 km distance, damaging the bottom of the lander with high-momentum rock impacts, and possibly tilting the lander as the excavated hole collapses to become a broad residual crater upon engine cutoff. Because of this, we deem that we will not have adequate safety margins to land humans on Mars unless we robotically stabilize the soil to form *in situ* landing pads prior to the mission. It will take a significant amount of time working in a harsh off-planet environment to develop and certify the new technologies and procedures for *in situ* landing pad construction. The only place to reasonably accomplish this is on the Moon.**

## Nomenclature

| | | |
|---|---|---|
| $a$ | = | acceleration |
| $C_D$ | = | coefficient of drag |
| $Kn$ | = | Knudsen number |
| $Re$ | = | Reynolds number |
| $v$ | = | velocity |
| $\Delta t$ | = | time step |

## I. Introduction

THIS paper will argue that we should not attempt human landings on Mars until we construct *in situ* martian landing pads, and that the surface of the Moon is the best place to develop the In Situ Resources Utilization (ISRU) technologies needed to do that. We formed this opinion not only from technical analysis (presented below), but also from seeing in prior human spaceflight programs how conservatism omitted at the beginning of a program is put back in later on. The spaceflight community is still in the very early stages of facing the reality of plume effects on Mars. We are sure that the community will eventually agree that we cannot land on Mars without some new and validated technologies to deal with the plume effects. We hope to accelerate this agreement, because the ideal place to develop and validate these technologies is on the surface of the Moon, and the time to begin developing them is now. Developing technology for Mars is of course one of the main reasons for returning to the Moon. President Bush stated in his speech laying out the *Vision for Space Exploration*, "We can use our time on the Moon to develop and test new approaches and technologies and systems that will allow us to function in other, more challenging environments…With the experience and knowledge gained on the Moon, we will then be ready to take the next steps of space exploration: human missions to Mars and to worlds beyond."[1]

---

[1] Physicist, Granular Mechanics and Regolith Operations Lab, Applied Sciences Division, KT-D3.
[2] Aerospace Engineer, Mission Analysis Group, Analex-20.
[3] Physicist, Advanced Systems, ASRC-15.
[4] Applications Scientist, Applied Science and Technology, ASRC-24.

NASA has extensive experience predicting and controlling the blast effects in terrestrial launch environments.[2] Even so, large particulates (rock- and boulder-sized) are sometimes liberated and ejected by the plume from the refractory concrete of terrestrial launch pads. This effect produces significant damage to the surrounding hardware (fences, lighting fixtures) with each Space Shuttle launch. Fig. 1 shows a portion of the damage, including the destruction of some fence sections at the launch pad perimeter, when over 3000 bricks were blown loose from the Space Shuttle flame trench with the launch of STS-124. When a rocket lands on the regolith of a planet or moon, we are at a disadvantage because the dust, soil, and rocks are much easier to liberate than on the specially-designed surfaces of terrestrial launch pads. The exhaust plume can blow the loose material at high velocity and this spray has the potential to damage the spacecraft and any surrounding hardware. It may also obscure the view of the landing zone, spoof sensors, contaminate surfaces and mechanisms with dust, and in some cases upset the soil so much that the lander is not stable or sufficiently upright after engine cutoff.

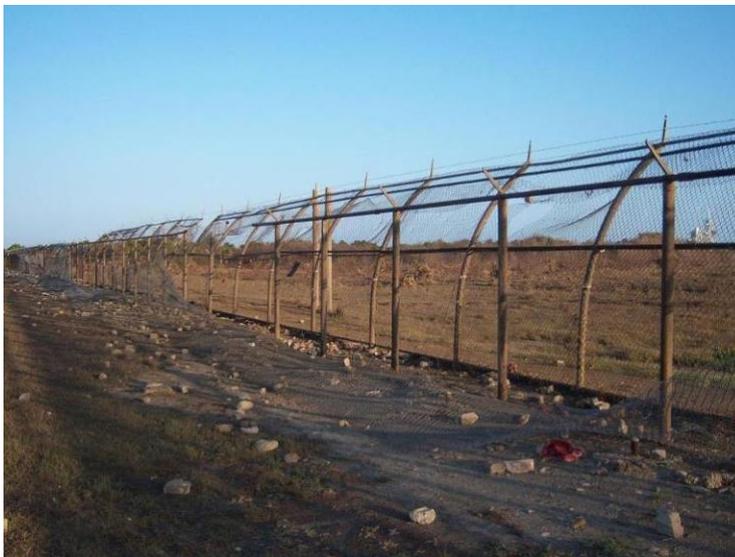

**Figure 1. Thousands of bricks blown loose by the STS-124 launch, destroying parts of the pad perimeter fence.**

The interaction of exhaust plumes with loose regolith material was first studied in the era leading up to the Apollo[3-11] and Viking[12-17] programs and again more recently.[18-26] There are several different flow regimes in the interaction between gas and granular media, and so the type of damage that could occur in a particular mission depends upon which flow regime is induced by the specifics of the propulsion system and the planetary environment. For example, on the Moon during most of the Apollo and Surveyor landings, there was generally no crater formation in the soil beneath the engines. Therefore, the relatively flat surface under the spacecraft ensured the ejecta would be blown horizontally away from the lander, and so the lander was safe from impacts. On the other hand, on Mars with the higher surface gravity and the very large landers needed for such long-distance human spaceflight, necessitating much greater thrust, and where the soil is softer (and desiccated at lower latitudes) and the planet's atmosphere dense enough to focus the engine exhaust onto a narrow patch of ground, very large craters will form beneath the engines (as we show below). This will cause a variety of problems not experienced during lunar exploration nor during the exploration of Mars with smaller robotic landers. For one thing, the narrowness of the initial crater will redirect the exhaust jet crater back upward, carrying entrained debris directly toward the landing spacecraft. It is a mistake to assume that results observed in the earlier programs such as Apollo and Viking are a direct indicator of what will occur in the future Moon and Mars programs where there will be different conditions and so different flow regimes will come into play. Specific analysis must be performed in each case.

These plume/soil interactions will affect the ISRU initiative in two ways. First, the spray of rocks and soil may damage ISRU hardware. Because the mission must make use of the ISRU-derived resources, the lander will be descending in the vicinity of the ISRU hardware by definition, putting it at risk. Second, the most effective way to mitigate the plume/soil problem may include robotic modification of the regolith at the landing site before the lander arrives, thus creating a spaceport. We might build a landing pad by cementation or other stabilizing technique so that the regolith resists erosion and cratering, and we might construct a berm to help block whatever material is liberated. This utilization of regolith constitutes ISRU by definition, and thus falls within the purview of the ISRU community. It may require the dual use of the ISRU excavator: when it is not collecting soil to extract chemical resources, it could build the berms, level the regolith in the landing zone, and serve as the mobile platform for a soil stabilization unit (e.g., microwave sintering unit). Thus, the plume/soil interactions are integral to the concerns of ISRU, both because the blast may damage ISRU hardware and because ISRU hardware may be essential in controlling the blast.

## II. Mechanisms of Gas Moving Soil

Research has identified four primary mechanisms by which gas moves soil. The first three were known during the Apollo era: Viscous Erosion (VE) as described by Bagnold,[27] Bearing Capacity Failure (BCF) as described by Alexander, et al.,[28] and Diffused Gas Eruption (DGE) as described by Scott and Ko.[29] More recently we have identified Diffusion-Driven Flow (DDF) as a distinct mechanism.[25] We describe these four mechanisms briefly below. We might also add mechanical shock to this list: that is, the mechanical shock induced by the initial overpressure of an igniting rocket engine. In experiments we see this overpressure send a shockwave into the soil, apparently compacting and dilating the soil as it passes. Shocks in granular media have been studied (see for example Ref. 30), but to our knowledge the shocks initiated by rocket engine ignition have not been studied.

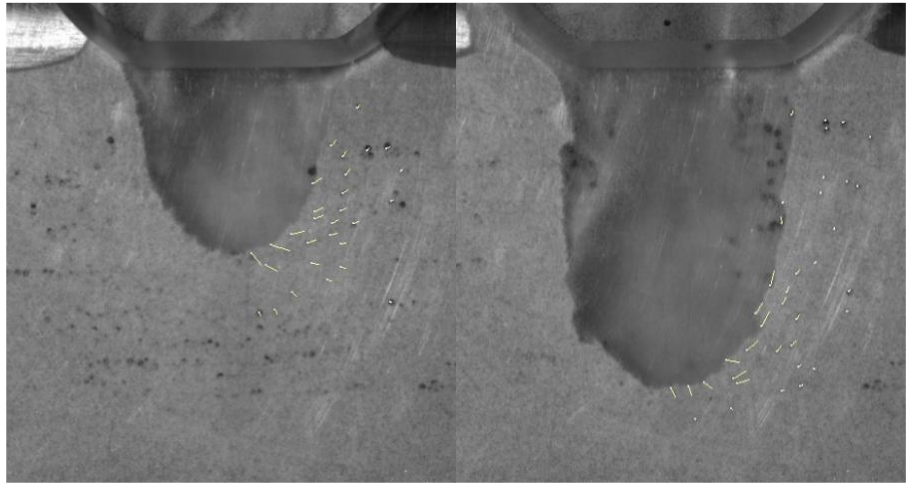

**Figure 2. Tracking individual sand grains beneath the craters.**

To describe these gas/soil interactions briefly, VE is the sweeping away of the top layer of grains by the tangential gas velocity above the soil. BCF is the bulk shearing of the soil to form a cup beneath a localized mechanical load applied to the surface, like a cone penetrometer pushing into the soil. In this case the mechanical loading is produced by the localized high pressure region of gas that develops under a perpendicularly impinging jet, i.e., the stagnation pressure. DGE is an auxiliary effect, which occurs when the stagnation pressure drives gas into the pores of the soil, only to erupt carrying soil with it at another location or time. DDF occurs when this same stagnation pressure of the jet drives gas through the soil so that the drag of the gas becomes a distributed body force within the soil, causing the soil to fail and shear in bulk. DDF and BCF both shear the soil in its bulk, but the former drives the soil tangentially to the crater's surface (radially from the tip and then up the sides), whereas BCF drives it perpendicularly to the crater's surface. In some cases, either DDF or BCF will dominate depending on whether the gas has sufficient time to diffuse through the soil relative to the time it takes the soil to shear. More generally, the gas diffuses only partially into the soil and the motion of the sand particles is intermediate to DDF and BCF,[25] moving at a diagonal to the surface of the crater. An example of this is shown in Fig. 2, where a gas jet is forming a crater against a window in the sandbox where the motion of the individual sand grains can be observed. In the left image the grains move diagonally to the crater's surface, indicating BCF and DDF are occurring in combination. In the right image most sand grains are moving tangential to the crater surface, indicating that the flow field of gas diffusing through the soil is fully developed so that only DDF is occurring.

The specific conditions of the planet and plume will govern which of these primary mechanisms are significant, and as a result it would be a mistake to directly compare the blast effects on the Earth, Moon or Mars with those that will occur on any other location where different mechanisms may come into play. For an example, Fig. 3 shows computational fluid dynamics (CFD) simulations using the Reynolds-Averaged Navier-Stokes equations for steady-state solutions of typical rocket exhaust plumes on the Earth, Mars, and the Moon. These were performed using the software package Fluent version 6.3.26 (Ansys, Inc., Lebanon, NH). Navier-Stokes equations break down in the limit of very rarefied gas, so the vacuum of the lunar case was approximated by setting the lowest background pressure that still permitted convergence of the solution, which turned out to be about 1 Pa. In some simulations we have systematically decreased the background pressure over several orders of magnitude and verified that only small changes occur near the stagnation region where the gas is dense, so the resulting solution is reasonably approximate in that region (adequate for our purposes while better methods are being developed). In both the terrestrial and martian cases, the jet is collimated by the surrounding atmosphere so that the shock above the soil and the stagnation region between the shock and the soil are limited to the diameter of the jet itself. In the lunar case the plume expands broadly into the (approximate) vacuum, resulting in a broad, bowl-shaped shockwave above the soil and a relatively broad stagnation region beneath the shock. Thus in the lunar case, the lower pressure and the more gradual pressure gradient around the stagnation region make BCF and DDF less likely to occur.

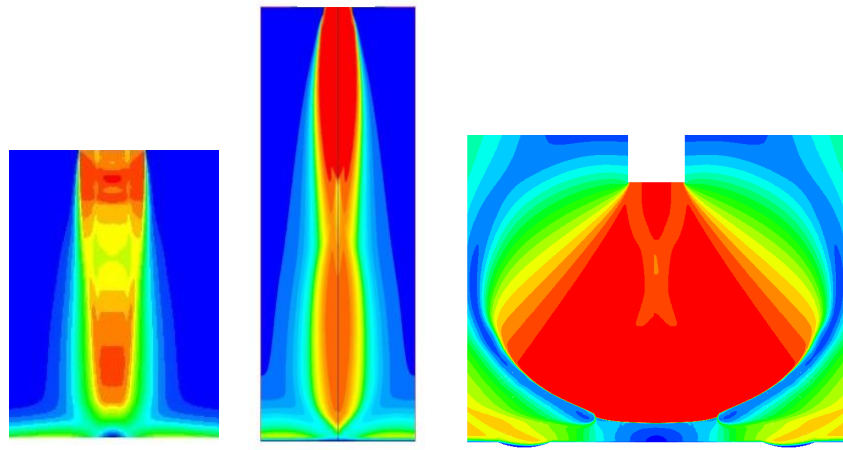

**Figure 3. Plume simulations for Earth (left), Mars (center), and the Moon (right). Red is the highest velocity and blue is the lowest.**

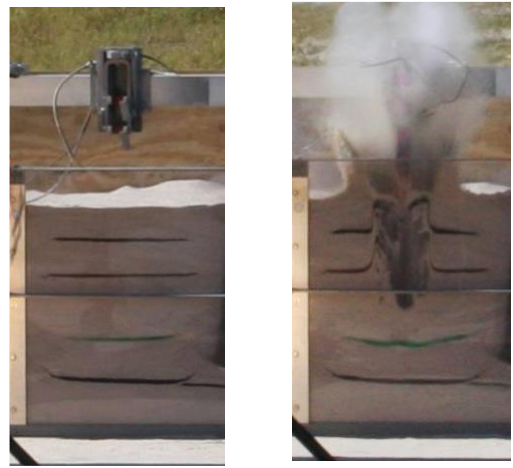

**Figure 4. Solid rocket motor firing in sand.**

## III. Terrestrial Plume Effects

Terrestrial experiments with subsonic jets in loose soil beds have demonstrated cases in which VE,[25,31] DGE,[29] BCF,[28] DDF,[25] or a combination of BCF/DDF[25] is the predominant phenomenon. BCF and DDF occur more readily when there is an ambient atmosphere to focus the plume, as illustrated in Fig. 3, creating a high stagnation pressure with steep pressure gradients around the edges. In terrestrial experiments with supersonic jets from solid rocket motors, such as in Fig. 4, BCF has been identified as the primary mechanism. In many tests, fluidization of a large quantity of soil follows the initial crater formation. For terrestrial launches on prepared concrete and brick launch pads, mechanical shock may be a significant mechanism to fracture and liberate surface material, which is then blown away by the plume. However, in the STS-124 Space Shuttle launch, it was the high velocity steady-state flow of gas through the flame trench, not the shock of the initial overpressure, that sucked thousands of bricks from the sides of the flame trench and ejected them over a wide area around the launch pad, as shown in Fig. 1. That is somewhat analogous to VE on a loose regolith, although on a much greater energy scale.

## IV. Lunar Plume Effects

During the Apollo program, NASA investigated the blowing of lunar soil by rocket exhaust plumes in order to ensure safe landings for the Lunar Modules. The investigations were primarily theoretical,[3] experimental,[4] or a combination of the two,[6] with very little numerical simulation because most numerical capability has been developed only in the years since then. After the lunar landings, the lack of visible craters under the landed spacecraft indicated that VE was the primary phenomenon. A high velocity flow of dust, sand, and possibly small gravel moves beneath the standoff shockwave of the plume in a nearly horizontal direction. The lifting of material appeared to be due to aerodynamic forces on the grains without much saltation due to the limited radial extent of the dense plume. The more violent BCF and DDF do not occur easily in the lunar environment because of the high mechanical strength of the unique lunar soil,[32] as well as its low permeability resisting gas diffusion,[33] and because the exhaust plumes expand widely in the lunar vacuum so that the stagnation pressure is not focused on a narrow patch of the soil. However, DGE may have occurred at least one time during the Surveyor V mission when the vernier engines were fired as a test after landing.[5] There were some large blasts of soil in the final moments of some landings (particularly Apollo 15),[22] which may have been due to enhanced viscous scouring at low altitude or the removal of soil mechanically fractured by the lander's contact probes. A plume-induced crater of approximately 444 liters volume was noted by the Apollo 14 crew.[19] It is possible that BCF or DDF may appear to some degree if the thrust of future lunar landers is larger than in the Apollo program; that is still an open question.

Even without the violent cratering phenomena, the lunar plume effects are very significant. The entrained dust particles obscured the view of the surface during landing, particularly on Apollo 12 and 15. The Apollo 12 mission report says, "On Apollo 12 the landing was essentially blind for approximately the last 40 feet."[34] In the crew debriefing, Pete Conrad reported,

...the dust went as far as I could see in any direction and completely obliterated craters and anything else. All I knew was there was ground underneath that dust. I had no problem with the dust, determining horizontal or lateral velocities, but I couldn't tell what was underneath me. I knew I was in a generally good area and I was just going to have to bite the bullet and land, because I couldn't tell whether there was a crater down there or not....[After landing] it turned out there were more craters around there than we realized, either because we didn't look before the dust started or because the dust obscured them.[35]

The spray of materials can also damage surrounding hardware on the Moon. For example, the Apollo 12 Lunar Module landed about 160 meters away from the deactivated Surveyor III spacecraft[9,10] as shown in Fig. 5. The Apollo astronauts returned portions of the Surveyor to the Earth for analysis, and it was discovered that permanent shadows were cast on its surface, because a thin layer of material had been scoured away by the sandblasting of entrained soil in the Lunar Module's plume. The surface of the Surveyor also had hundreds of pits (micro-craters on the order of 100 microns in diameter) from the impact of high-velocity soil particles,[9] and a recessed location on the Surveyor's camera was contaminated with lunar fines (up to ~150 micron particles) where they had been blown through an inspection hole. It was determined that the velocity of particles striking the Surveyor was between 300 m/s and 2 km/s.[36] At the time, the Surveyor spacecraft was already deactivated when the Lunar Module landed. For functional hardware on the Moon, this sort of treatment will not be acceptable. The scouring effects of the spray may ruin surface coatings, reflective blankets, and optics, and the injection of dust into mechanical joints may cause increased friction, jamming and mechanical wear. Recent experiments blowing simulated lunar soil at spacecraft materials has shown that even at much slower velocities the abrasiveness of the particulates causes unacceptable damage.[37] In the future, when spacecraft return to the same location multiple times in order to build a lunar base or to perform maintenance on scientific hardware, it will be necessary to prevent or block the spray.

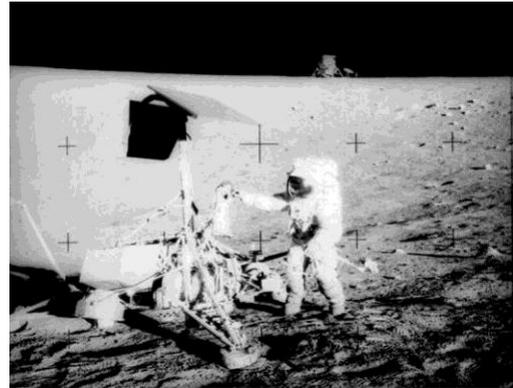

**Figure 5. Surveyor III spacecraft with Apollo 12 Lunar Module on horizon.**

Until now, the best predictions for lunar soil erosion have been direct measurements taken from the Apollo landings. The morphology of the blowing dust blanket can be measured by the shapes of the Lunar Module shadows that drape across it. Analysis of these shadows in the landing videos shows that the dust is blowing at an approximately 1 to 3 degree elevation angle in all missions (but Apollo 15, discussed below) away from the plume's stagnation region.[22] Numerical simulation[23] of the plumes indicates the dust velocity is on the order of 1 km/s, in agreement with Surveyor III data, and blowing at angles between 1 and 3 degrees, in good agreement with the Apollo landing videos.

Because the ejecta blows at such low angles, it may be economical to build a berm around the landing site to block the majority of this spray.[38] A berm is built from *in situ* materials, requiring no mass to be lifted from Earth apart from the machinery that builds it. An excavator designed for oxygen production on the Moon may serve the dual-purpose to build the berm. Another option is to place inflatable screens[39] around the landing site or in front of critical hardware. This has a mass-per-linear-distance cost, but may be quick and easy to assemble and thus may have a reliability advantage compared to the long-duration usage of excavators to build berms. Trade studies are needed to decide.

In addition to blocking the ejecta, it may be necessary to level and stabilize the regolith in the landing zone via sintering, cementation, application of palliatives, or the use of a fabric mat. This is because the ejecta blows at a low angle only when the surface is flat and remains flat. Ejecta blowing out of non-flat surface features will ramped upward at higher angles. This is seen in the Apollo landing videos, in which streaks of dust emanate from meter-scale craters at angles that are not a constant 3 degrees and are modulated by the spacecraft thrust.[22] Also, in Apollo 15, the ejection angle was measured to be between 7 and 11 degrees due to the lander descending over a broad, shallow crater that ramped the dust.[22] Furthermore, in Apollo 15 a dense blast of soil was seen in the final seconds of landing that sent ejecta to some angle higher than 22 degrees.[22] This high angle was presumably caused by the engine nozzle coming very close to the loose soil of the crater rim, blowing out a large quantity of soil and thus forming for a moment a small depression under the nozzle that ramped the material into even higher angles. As mentioned above, a crater was also identified by the crew of Apollo 14; it was lying along the landing trajectory just behind the final resting position of the nozzle and having the appearance of being made by the plume.[19] So in general it seems the Apollo plume conditions were just borderline to being able to cause small-to-medium sized craters or scour holes, and the larger thrust of the Altair in the upcoming return to the Moon may be adequate to cause larger craters or scour holes. If a scour hole forms, then it will ramp the ejecta to higher angles and then the larger, slower ejecta will rain down upon the outpost, rendering berms or screens ineffective. So a complete solution to the pluming problems on the Moon probably requires surface stabilization.

If the surface is stabilized, then it is a valid question whether berms are required, too. Conservatism would argue "yes" because failure of the stabilized surface would create larger, possibly platy-shaped ejecta with large surface-to-mass ratio, which would be blown more easily than the natural rocks and thus cause greater damage to the outpost. Also, it is simply good practice to put physical barriers between people or hardware assets and rocket exhaust, and furthermore the chemical contamination of the plume must be prevented from contaminating surfaces all over the outpost. Trade studies are needed to definitively answer this question.

In addition to these countermeasures, we should consider what technology to develop on the Moon purely because we will need it for Mars. This may also play a role in determining what technologies we use on the Moon. A plume solution that is not optimum for the Moon may still be the best choice for the Moon because of the benefit it gives us in reaching Mars.

## V. Martian Plume Effects

Prior to the Viking landings on Mars, it was determined that BCF would occur and produce scientifically unacceptable disturbance in the soil sampling area.[15,16] The spacecraft engines were re-designed so each one's nozzle was replaced by a "showerhead" design with 18 tiny nozzles. This enhanced the turbulent mixing of the exhaust jets with the ambient atmosphere so that the jets were extincted at a shorter range. Nonetheless, shallow craters were confirmed to occur under the engines within view of the cameras, and at the Viking Lander 2 site there was a cluster of small craters corresponding to each of the individual nozzles.[17] In the recent Phoenix mission, the lander descent thrusters cleared away patches of soil exposing the ice table several centimeters below the surface. The presence of the shallow ice table masked the possibility that the thrusters might induce deep cratering in the soil. The Mars Science Lab has been designed with a "sky crane" system to lower the rover to the regolith on cables while the descent stage with its thrusters remains high above the ground. Our recent work (reported here) shows that when an even larger 40 MT spacecraft (84 times the mass of the Mars Science Lab) lands on the regolith of Mars, the blast effects will be significantly greater than our prior experiences in Apollo and Viking. There will be fluidization and cratering of the soil, which may push deep beneath the surface (at least when landing in mid-to-low latitudes where there is a deep bed of soil overlying the ice table). We address three major concerns: (1) regolith material will be ejected into high trajectories and may damage surrounding hardware far away; (2) rocks will strike the landing spacecraft itself at high velocity, and (3) the fluidized soil will collapse into a broad crater after engines are cut off, possibly upsetting the stability of the landed spacecraft.

### A. Martian Conditions

On Mars the atmospheric density and the gravity are both intermediate to the lunar and terrestrial cases. The Martian atmosphere is sufficiently rarefied that the Knudsen number for small dust-sized particles will be on the order of unity and thus the gas flow around these particles will not be well-described by the Navier-Stokes equations. It will be necessary to account for this rarefaction in calculating the coefficient of drag for smaller particles. The carbon dioxide atmosphere has significant bulk viscosity, which is negligible in the terrestrial atmosphere, but to-date the effects of bulk viscosity for martian ballistics has not been well characterized and is neglected in this study. It is expected that bulk viscosity will only reduce the distance that particulates will travel, and so if bulk viscosity is significant then this study represents a worse case.

The atmosphere on Mars will collimate the engine plumes so that the interaction with soil will be much closer to the terrestrial case (experiments in sand beds in an atmosphere) than the lunar case (exhaust plume expanding into a vacuum). The standoff shock over the soil will be only as wide as the collimated plume with an abrupt drop in stagnation pressure around its circumference, and this will be far more likely to penetrate the soil and create the type of deep craters seen in terrestrial testing. Hence, the effects on Mars will be dramatically different than on the Moon.

The regolith on Mars[40-42] is probably very diverse but appears to be more akin to terrestrial than lunar regolith[32] because it is looser, less cohesive, more porous, and overall more easily excavated by an impinging jet than the lunar regolith. On the Moon in the absence of an atmosphere to regulate temperature, the extreme monthly thermal cycling of the regolith produces strong lunar quakes in the crust and over geologic time these have shaken down and compacted the soil to a very dense state. On Mars as on Earth, the temperature of the soil is moderated by the atmosphere so that heat is not directly radiated away to space, and the much shorter diurnal period of the thermal cycling, so the soil will not be as compacted by thermal cycling and the resultant shaking as on the Moon. Further, the active aeolian processes might have contributed to its relatively loose state at the surface, and its prior hydrological processes may have contributed to geological sorting of the particle sizes in some regions, enhancing the permeability and porosity of the soil. The particle shapes of the weathered martian soil are probably more rounded like terrestrial particles than the sharp, interlocking shapes of the un-weathered lunar particles. All these factors make the soil much weaker on Mars than on the Moon and so cratering effects are expected to be much greater for martian landings.

The mass and thrust of launching and landing spacecraft on Mars will also be intermediate to the conditions of the Moon and Earth. Mars landers will certainly have less mass and thrust than the Space Shuttle and could do less damage to a brick-lined flame trench. However, with its larger crew and longer surface stay than in the Apollo missions, it will have a much greater landed mass than the Apollo Lunar Modules. Multiplying that greater mass by the greater surface gravity of Mars relative to the Moon, and the thrust to land these spacecraft must be much greater than in the lunar case.

### B. Deep cratering on Mars

To characterize the martian plume effects, the principal question is whether or not the deep cratering of BCF and/or DDF will occur, producing the deep, narrow crater characteristic of those processes. If so, then the gas exiting the crater will eject rocks and soil into the vertical direction as shown in Fig. 6, and then, as the crater erodes and broadens, into a wide range of elevation angles. There are presently two methods to predict the onset of deep cratering and both predict that deep cratering will occur for human-sized landers on Mars.

The first method is the purely empirical limit of 0.3 kPa stagnation pressure on the soil, determined during the Viking tests.[16] While the Viking landers were successfully kept below this limit, the heavy human-sized landers must necessarily exceed it. For a 40 MT lander with four engines using nominal engine conditions, the pressure is calculated to be 47 kPa, more than 150 times the limit. For a 60 MT lander this increases to 71 kPa, about 240 times the limit. This implies that deep cratering will undoubtedly occur. However, this empirical limit does not take into account the details of the physics, and we cannot be sure that the gas/soil flow regime present in the Viking tests will be the same that occurs with landers having significantly more thrust.

The second method to predict the deep cratering is an analytical theory developed by Alexander, et al.[28] To date, this is the only theory or model that attempts to predict the onset and depth of cratering and so it is worth using for comparison with prior work. Nevertheless, it is necessary to point out that it contains three known flaws and so its predictions may not be valid. First, it assumes that the stagnant gas under the impinging plume can play two roles simultaneously: (1) applying mechanical pressure at the surface of the soil, and (2) diffusing fully into the soil to reach a steady-state flow condition that maximally weakens the soil as in Terzaghi's effective stress hypothesis. This is wrong because the net mechanical pressure applied to the surface of the soil vanishes as the gas reaches steady-state diffusion into the soil (i.e., the difference in pressure between the top and bottom sides of the uppermost layer of sand grains becomes nearly zero so there is negligible mechanical loading applied to the surface). Thus, in the theory the surface pressure driving the bearing capacity failure should have been reduced to a degree commensurate with the diffusion. Second, Terzaghi's effective stress hypothesis deals with essentially static fluid in the soil, the pressure of which relieves some of the hydrostatic portion of the stress in the mechanical skeleton of the soil. However, in this case the fluid is highly dynamic and its aerodynamic drag through the soil produces a distributed body force that actually contributes to the stress in the mechanical skeleton of the soil rather than relieving it. Thus, the pressure at the surface of the soil is reduced as diffusion proceeds, and it is replaced by a body drag force throughout the bulk of the soil, and this transitions the cratering effects from a case of BCF to a case of DDF. In the latter the soil moves in a direction that is perpendicular to the former, and thus the Alexander, et al., theory is not adequate to describe both cases. Unfortunately, it is difficult to write equations for DDF as analytically tractable as the equations written by Alexander, et al for BCF. Third, the Alexander, et al, theory assumes a geometry for soil motion that is unrealistically restricted to a narrow cylinder around the crater, resulting in too much predicted soil resistance. Recent experiments with a jets impinging on sand against a glass window have shown that the shearing geometry is much wider than this. In summary, the Alexander, et al., theory omitting gas diffusion will under-predict the cratering because it neglects both DDF and the lower-resistance shearing geometry of the soil, but the theory including gas diffusion and Terzaghi's effective stress hypothesis is conceptually wrong and it is unknown whether it will over- or under-predict the cratering.

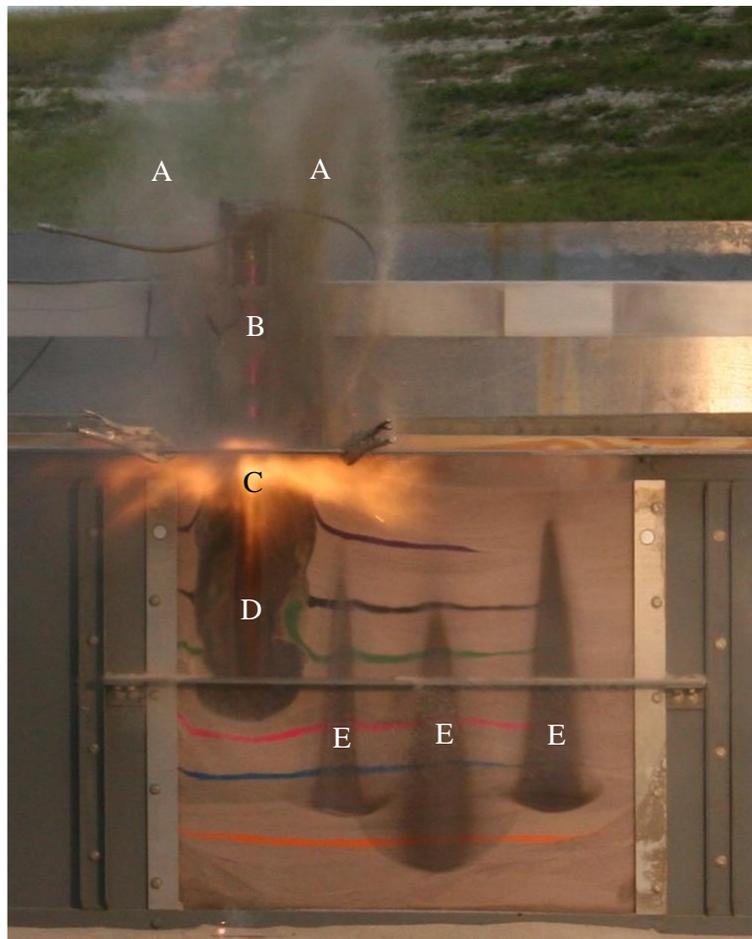

**Figure 6. Test with 50 N solid rocket motor firing into a 0.77 m deep sand bed interspersed with colored layers of sand.**

With this in mind, the theory has been applied as follows. In the Viking era, Ko applied the theory including gas diffusion for the original design of a Viking lander engine (prior to the multi-nozzle engine) in martian conditions and predicted that BCF would occur from a descent height of 5 meters down, producing a crater as deep as 60 cm.[15] A much larger engine for a 40 MT lander would therefore be expected nominally to produce a deeper crater. The present study has performed the calculations for such a case, but omitting the gas diffusion in order to obtain an assuredly low-end prediction thus (at least) bounding the problem. The calculations were performed using soil cohesion and friction angle inferred from recent data collected at Meridiani Planum by the Opportunity rover.[42] The calculations predict that the crater will be 2.0 m deep. (This is coincidentally the approximate, inferred depth to martian bedrock in the Meridiani Planum region.[40]) A sensitivity study was performed on the various parameters in the theory and it was shown that under all conceivable conditions this cratering will occur. To demonstrate that this 2.0 m prediction is an under-prediction of the depth of the crater, the theory was also applied to the recent terrestrial test conditions of 50 N solid rocket motors firing into a deep quartz sand bed, and the theory without gas diffusion predicts only 5 cm deep crater compared with the actual craters that repeatedly formed to 55 cm depth. Thus, the theory without gas diffusion grossly under-predicts by a factor of 11 in these test conditions. If the theory were updated to properly include gas diffusion leading to DDF, and also to use a more realistic shearing geometry, then the prediction for a 40 MT lander would be undoubtedly be deeper than 2.0 m, although we cannot presently say how deep.

Integrated numerical models of plume/soil interactions for continuum through rarefied conditions are being developed under NASA contracts, and they should be capable of predicting all these effects. In the meantime, we can conclude with reasonable confidence that narrow cratering deeper than 2 m will occur beneath the large landers needed for human exploration. Simulation of the ballistics will therefore assume typical conditions for gas exiting a narrow crater as the worst situation that will occur during a nominal landing. As the crater grows and widens (as demonstrated in tests), then the exiting gas velocities will lessen. The ballistics calculated from the narrow exit condition will therefore represent worst-case.

## C. Martian ballistics compared to terrestrial ballistics

The following analysis shows that the plume will blow dust, sand, gravel, and even rocks up to about 7 cm in diameter at high velocity. These ejecta will cause significant damage to any hardware that is already placed on the martian surface within the blast radius. Ballistics equations were developed for the martian environment, accounting for aerodynamic drag and gravity. For aerodynamic drag on a particle neglecting rarefaction we have used the Schiller and Nauman[43] correlation

$$C_D = \left(\frac{24}{\text{Re}}\right)\left(1 + 0.15 \text{Re}^{0.687}\right) \qquad (1)$$

where the Reynolds number Re is based on a particle diameter. To account for rarefaction, the following factor by Carlson and Hoglund[44] is multiplied onto the coefficient of drag:

$$\frac{1}{1+ \text{Kn}\left[3.82 + 1.28\exp\left(-\text{Kn}^{-1}\right)\right]} \quad (2)$$

where the Knudsen number Kn is based on a particle diameter. This method appears to underpredict the drag on the smallest particles at high Reynolds and Knudsen numbers, but these small particles have low terminal velocity and low mass and so are not a concern for damage calculations. The trajectory was integrated using the simple Euler's method, but with the adaptive time-step $\Delta t = v/ak$ where $v$ and $a$ were the particle's velocity and acceleration in the prior time-step, and $k$ is an arbitrary factor that remains constant throughout the trajectory. In successive simulations of the same set of conditions, $k$ was systematically increased to as high as 1000, verifying that further increases no longer affected the trajectory and thus the time step was sufficiently small that the accumulation of integration errors from Euler's method was negligible. The algorithms were implemented in Mathematica 6.0.1 (Wolfram Research, Inc., Champaign, IL).

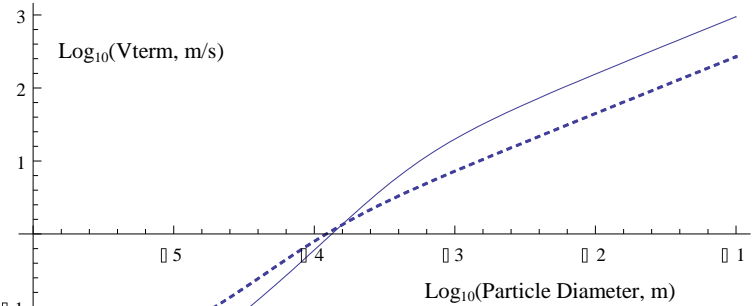

Figure 7. Terminal velocity as a function of particle size for Earth (dashed) and Mars (solid).

To develop a "sense" for martian ballistics, this code was first applied to particles free-falling in both martian and terrestrial atmospheres to compare the terminal velocities in each case. The results in Fig. 7 indicate that large silica particles will fall 3.5 times faster on Mars than on Earth, but small silica particles will fall 3.15 times faster on Earth than on Mars. The transitional regime is in the range from 10 microns to 1 mm, with particles sized 157 microns (fine sand) having the exact same terminal velocity of 1.33 m/s on either planet. Of course, these martian calculations will vary seasonally and with altitude. The calculations here used an atmospheric temperature of 193 K and a pressure of 6.2 mbar for specificity.

**D. Plume and crater model**

The martian atmosphere is sufficiently dense that standard computational fluid dynamics software can accurately model the plume. A number of plume simulations have been performed using realistic nozzles and engine conditions, and an example is shown in Fig. 8. The ratio of the engine nozzle diameter to crater diameter in this example was chosen as typical from terrestrial experiments. With the geometry selected here the engine exhaust gas exits the crater in a jet around the circumference of the crater, coaxial to the engine exhaust, and with maximum velocity near the lip where entrainment of particulates occurs. This agrees with test results, in which a vertical jet of entrained material leaves the crater, shown in Fig. 6. In other simulations we used crater geometries to represent the widening of the top of the crater that will occur through erosion, as seen experimentally.

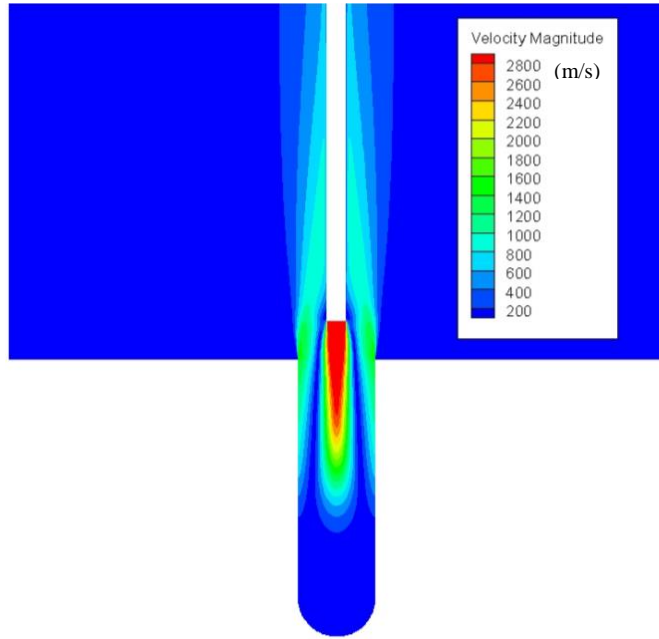

Figure 8. CFD simulation of martian exhaust plume into narrow crater, showing coaxial jet exiting from crater.

Since the cloud of small particulates blown from the crater will be stopped by drag forces at ranges close to the lander, the radius of the blast zone will be determined by the larger particles (coarse sand, gravel, or rocks) that travel the farthest. This damaging material will fall sparsely throughout the large blast zone and the probability of being hit will depend on the quantity of each particle size that is ejected into each trajectory angle. Since even one unacceptably high impact of large material must be prevented, we can define the blast radius simply as the maximum distance that any unacceptable impacts could occur. Our goal is therefore to identify the maximum distance any particle can travel and the maximum impact that could possibly occur at each distance.

As the ballistics code integrates the trajectories using Euler's method, it could calculate the drag force by interpolating the gas conditions in the CFD solution to the particle's exact position at each time step. However, we do not believe this would be the best method at the present. For one thing, the resulting blast radius would be no better than a crude order-of-magnitude estimate, because the CFD simulations were performed using only rough guesses of the crater diameters and geometries (albeit based on experimental observations), and therefore are not really accurate. The flow codes to predict actual crater evolution are not yet available. So rather than performing this precise interpolation through the CFD solutions, we could do just as well using a crude estimation of the drag. For another thing, the CFD simulations we have performed might not include the worst-case conditions that could occur out of the family of all possible crater shapes. We found from the CFD simulations that the fastest gas velocity (about 1300 m/s for a nominal 40 MT lander) occurs when the crater is narrowest, ejecting the particles straight up. But that is not the direction the produces the farthest blast radius. As the crater widens and thus sprays materials outward and further away, the gas velocity exiting the crater is generally slower. We do not know what crater shape produces the worst-case combination of gas speed and ejection angle. Furthermore, real craters are dynamic, chaotic, and partially filled with fluidized soil, so the idealized CFD simulations are not realistic. We can bracket (or approximate) the longest trajectory if we use the fastest-case (or a reasonable) gas velocity and arbitrarily set its direction through the full range from horizontal to vertical. Therefore, we wrote the ballistics code to use a constant 1000 m/s gas velocity exiting the crater at any angle selected by the user. The particle is accelerated by the vectorial sum of gravity and the drag force resulting from the relative velocity of the particle in this gas. The code applies this across a reasonable distance (5 m based on the CFD simulations) then switches abruptly to calculating the drag force of an ambient atmosphere for the remainder of the trajectory. The resulting blast radius calculated by this method will be to the correct order-of-magnitude and this estimate is adequate for now, until the complete physics-based soil/gas flow codes have been developed.

**E. Blast zone predictions**

Particle trajectories were obtained in this approximation for the full range of particle sizes and gas velocity angles. An example of some typical trajectories are shown in Fig. 9. Particles in the 3 mm size range (coarse sand to fine gravel) traveled the farthest (700 m) and hit with the highest velocity (43 m/s). The heavier particles (as shown with the symmetrically arched trajectories) do not travel as far because the plume is not able to accelerate them to as high an initial velocity as the 3mm particles. The lighter particles (as shown with the very asymmetric trajectories) have initially greater velocity leaving the plume but do not travel as far because they lose kinetic energy to the ambient atmosphere and then fall straight down at their terminal velocities. Hardware designers might find that the momentum and energy of these particles is not too severe and may be successfully shielded. Therefore, it may be possible to set a smaller blast radius. Although the 3 mm particles have the highest impact velocity, the heavier particles falling at smaller radii have greater impact momentum and energies as shown in Figs. 10 and 11, respectively. The variety of impact momenta and energies that can occur at each distance are shown in Figs. 12 and 13, respectively. Shielding the hardware to withstand these impacts will require mass, thus reducing the useful mass available to do other functions.

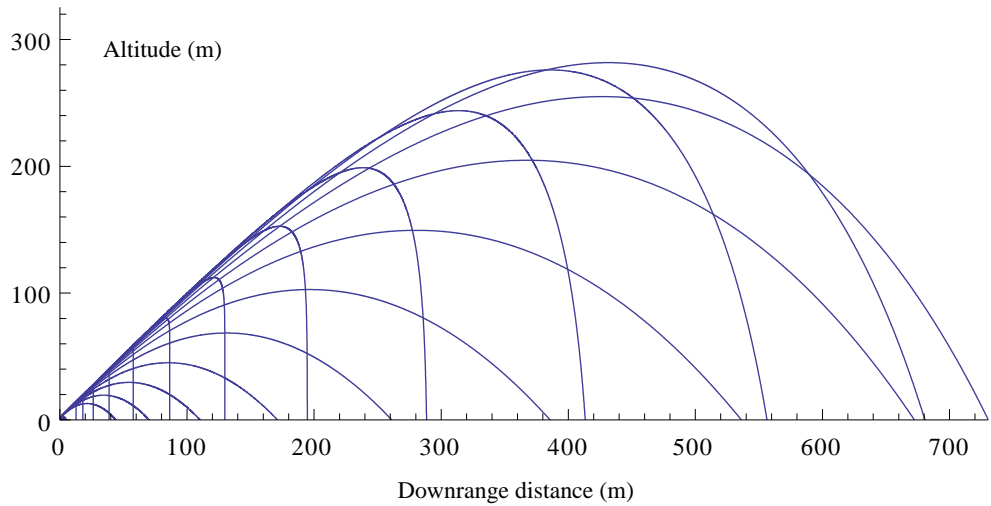

**Figure 9. Particle trajectories ejected from crater at 45 degrees for 23 different particle sizes from 50 microns to 10 cm. Particles in the 3mm size range travel the farthest and hit with the highest velocity.**

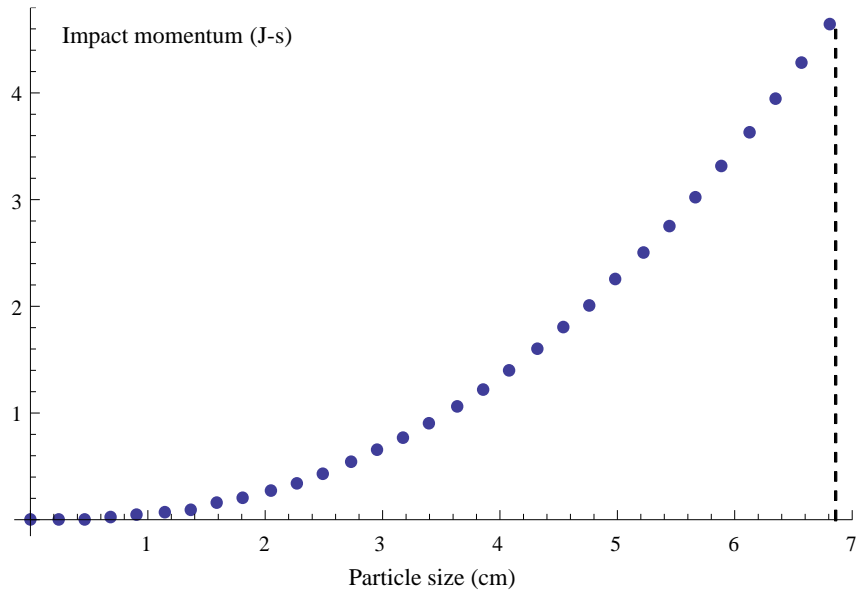

**Figure 10. Impact momentum versus particle size. Dashed vertical line is limit of largest particle that can be ejected by the plume**

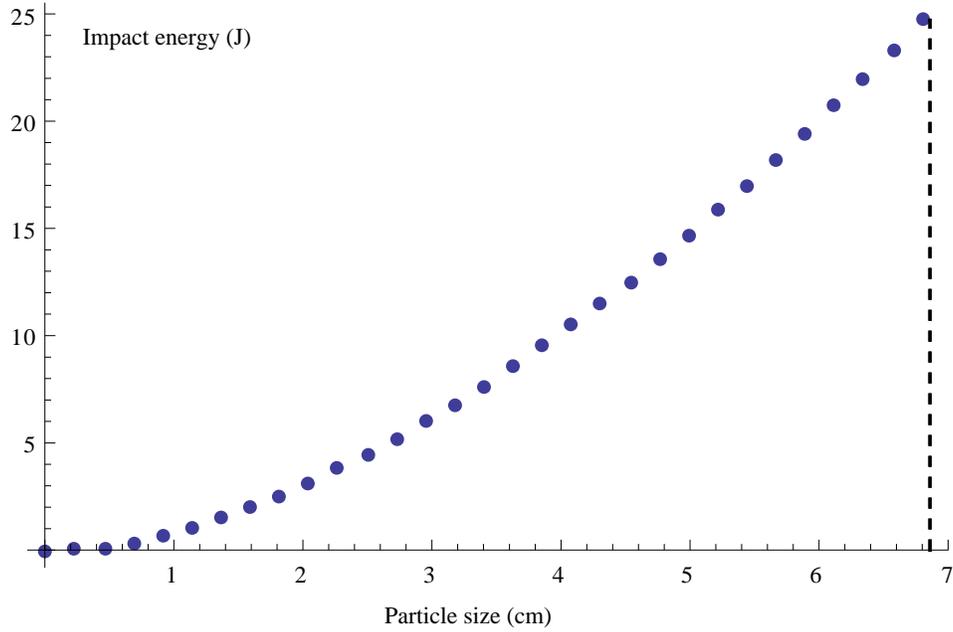

**Figure 11. Impact energy versus particle size. Dashed vertical line is limit of largest particle that can be ejected by the plume**

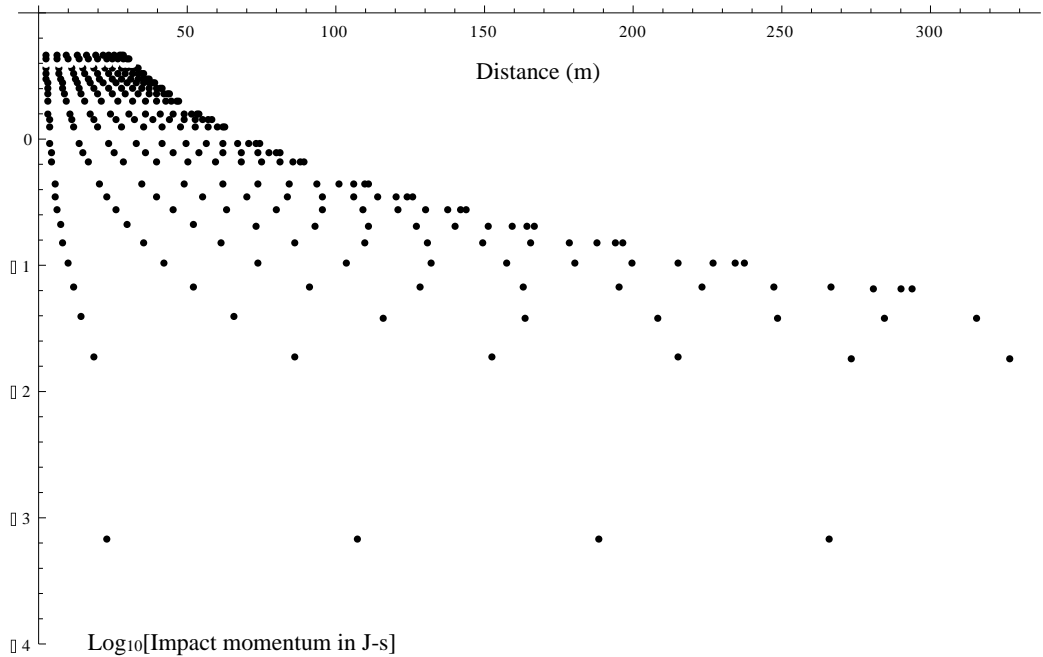

**Figure 12. Scatterplot of impact momentum versus impact distance for a sampling of particle sizes and ejection angles. Hardware specification of allowable impact momenta may identify a safe landing distance from this plot.**

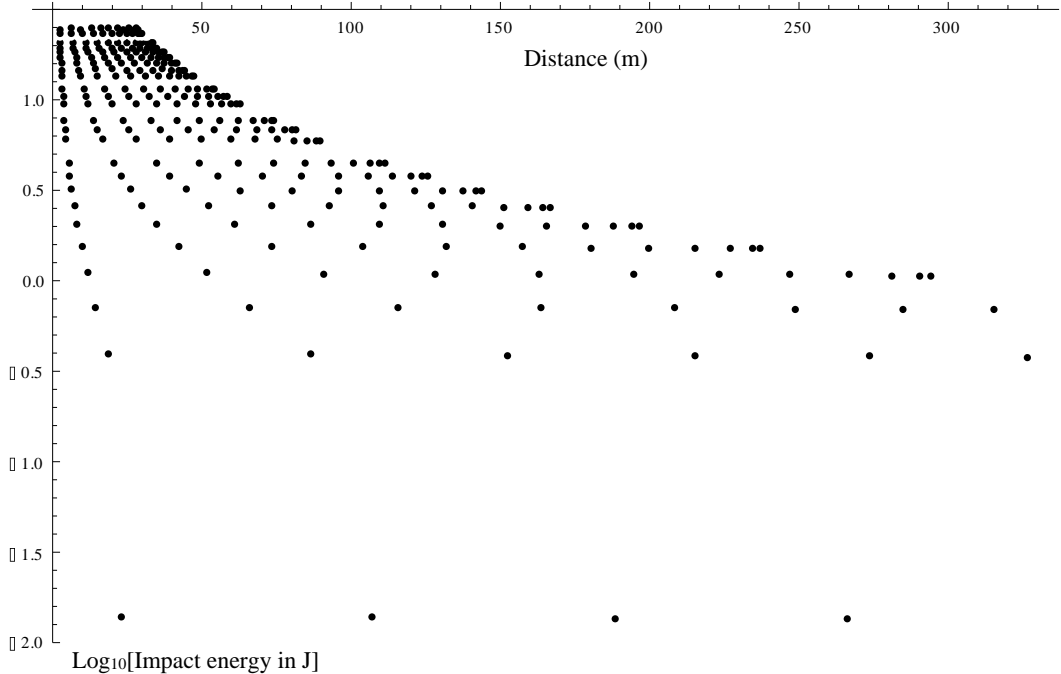

**Figure 13. Scatterplot of impact energy versus impact distance for a sampling of particle sizes and ejection angles. Hardware specification of allowable impact energies may identify a safe landing distance from this plot.**

**F. Ejecta impingements on the lander**

Ejecta will not only strike hardware in the surrounding area, but will strike the bottom of the lander itself. This is because a narrow crater ejects material vertically. As time goes on (and the lander descends), the crater is widened by erosion so we might expect that all vertical ejection occurs while the lander is still very high. However, in some tests we have seen the craters widen and then become narrow again as fluidized soil collapses into them. Also, as the lander nears the surface, the increased focusing of the plume on the regolith may lead to a renewed burst of BCF/DDF, forming a new narrow crater at the bottom of the existing wider crater. In general, experiments show that the cratering effects become increasingly chaotic as time passes and ejection of materials can be in any direction at any time.

Impacts to the lander were estimated using the same ballistics code described above but with the ejection angle selected to be vertical. The particles then travel through the ambient atmosphere that exists above the crater and adjacent to the jet as shown in Fig. 8. They either reach a maximum height and begin falling or they strike the base of the lander. Our CFD simulations modeled the engine nozzle but not the base of the lander just above the nozzle, so we have not characterized the effect of the base pressure that will build up when it is near the ground. Impact velocities, momenta, and energies as a function of particle size are shown in Figs. 14, 15, and 16, respectively. Generally, these increase as the lander descends to lower altitudes. Shielding to protect the bottom of the lander or ruggedization sufficient to withstand these momenta and energies will be required. For example, the nozzles must be protected, since ejecta blown by one engine may strike the nozzle of another engine. This will add significant mass to the lander, thus reducing payload mass.

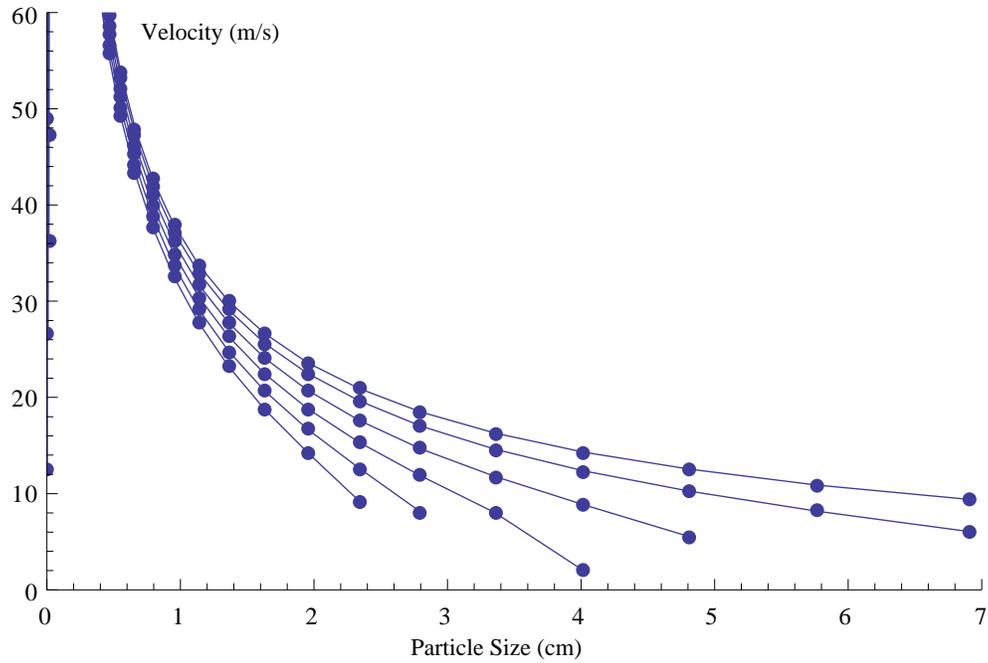

**Figure 14. Impact velocity on the bottom of the lander as a function of particle size for several lander heights: 2 m (top curve), 10 m, 20 m, 30 m, and 50 m (bottom curve).**

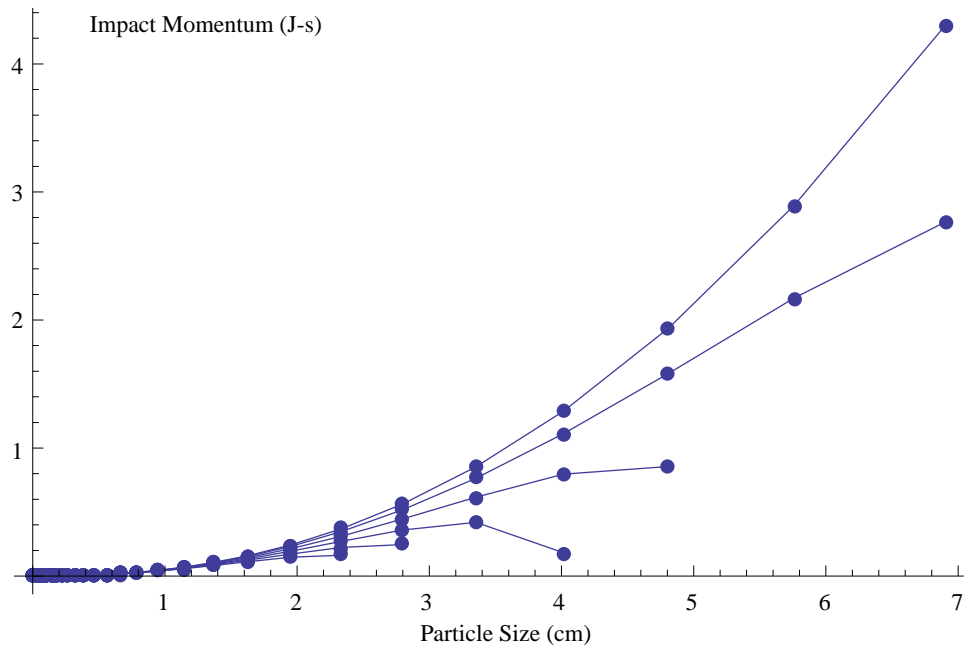

**Figure 15. Impact momentum on the bottom of the lander as a function of particle size for several lander heights: 2 m (top curve), 10 m, 20 m, 30 m, and 50 m (bottom curve)**

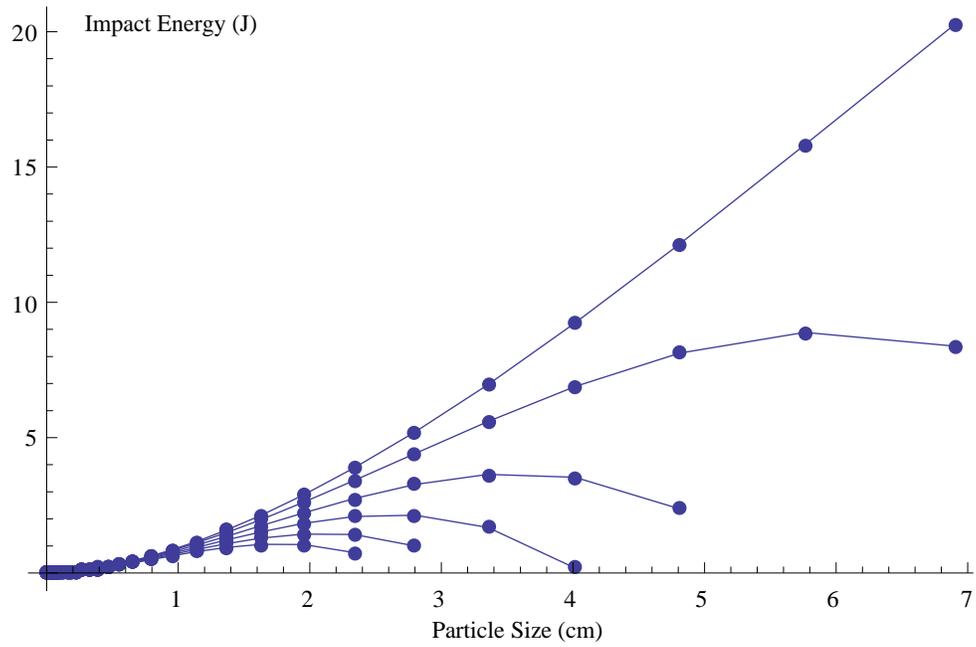

**Figure 16. Impact energy on the bottom of the lander as a function of particle size for several lander heights: 2 m (top curve), 10 m, 20 m, 30 m, and 50 m (bottom curve)**

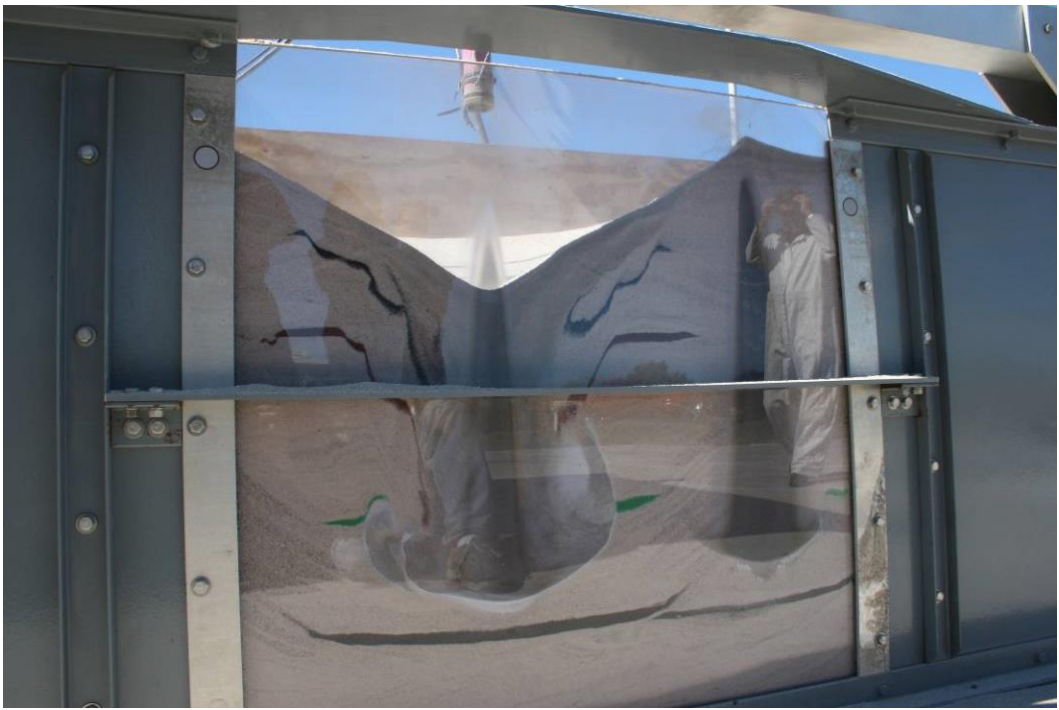

**Figure 17. Residual crater after solid motor firing is complete. The original crater depth before collapse is witnessed by the deep scorch on the glass. The second scorch (on the right) is from an earlier test before the sandbox was refilled.**

**G. Regolith stability after landing**

In addition to the concerns with plume debris striking the surrounding hardware or the lander itself, another concern is with the stability of the soil after the engines shut off. Experiments have shown that the deep, narrow crater beneath the plume will maintain its shape and slowly broaden as long as the engine continues firing, but when the engine cuts off then the crater collapses into a cone-shaped "residual crater" as shown in Fig. 17. We have measured the volume of the deep cylindrical crater just prior to engine cutoff and also the volume of the residual crater after engine cutoff and have verified that they (trivially) equal one another. So if we can estimate the volume of the cylindrical crater in a Mars landing, then this will also be the estimate for the residual crater at engine cutoff. With knowledge of the crater's inner slope we can then calculate a depth and a diameter. For non-cohesive or only slightly cohesive soils, such as the upper layers of soil measured by the Mars Exploration Rovers, the angle of repose is about 20 degrees. We assume for order-of-magnitude that the initial crater is of the diameter shown in Fig. 8, and has a depth of only 2 meters to martian bedrock (or the ice table) with very little additional erosion to widen it after reaching bedrock. A cone of 20 degrees with this volume has a width of 7.3 meters. With four engines clustered toward the center of the lander, the single residual crater resulting from the collapse of four closely spaced holes will have a width of 11.6 m. Thus, the residual crater will be very broad and may extend well past the lander's footpads. Cohesion in the soil due to ices or chemical cementation may keep the hole from collapsing immediately, or from collapsing uniformly all around. The ground beneath the lander would be unsafe because it could collapse at any moment as ice sublimates from the hole or as soil creep under the load of the spacecraft eventually causes the cohesion to fail.

One way to mitigate this is to increase the width of the landing gear to extend beyond the zone of unstable soil. However, this wider landing gear will significantly increase the lander mass and therefore significantly decrease the mass of the payload. Also, if the bedrock or ice table is lower and the crater is excavated more deeply than 2 meters, or if the erosion removes significantly more soil by widening the initial crater, then the crater will have greater volume and the residual crater may still reach to the footpads. The above calculations have significantly under-predicted the cratering effects and so the crater may actually be wider, but we are not able to make more accurate predictions at the present. Not only does collapsing soil threaten to tilt the spacecraft, but if bare bedrock or boulders are exposed, then that might present an unexpectedly uneven landing surface. In any case, there is a real possibility that the lander could be unleveled after engine shutoff and it will be difficult to convince ourselves in an atmosphere of conservatism that we have positive safety margins for a human-tended mission. Only a moderate tilt of the lander due to soil collapsing could make it impossible for the ascent module to safely separate and bring the crew home from Mars.

## VI. Mitigation and ISRU Implications

Adding shielding to the bottom of the spacecraft and increasing the length of the landing gear both add significant mass to the lander. Even accepting the significant loss of payload mass, this approach may not provide the assured safety margins that a human-tended mission to Mars would require and eventually get as conservatism and reality settled into the program. Fortunately, there is another option that has the potential to both provide clear safety margins and preserve payload mass. A landing site could be prepared robotically prior to human arrival, and then it would be possible to use smaller landing gear and omit the extra blast shielding. The robotic site-preparation hardware could be sent to Mars on the first flight opportunity. It would then begin preparing two landing sites. The first of these would be used by an un-fueled ascent lander (to be fueled on Mar by ISRU) that would be sent on the second flight opportunity. The other landing site would be for the humans' arrival on the third flight opportunity. Variations on this idea can be devised to decrease the number of required flight opportunities from three to two, depending on what one is willing to give up. For example, the un-fueled ascent lander could be landed without a prepared surface on the first flight opportunity, along with the robotic hardware, because the human mission could be simply called off if the cratering or ejecta impacts caused its landing to be unsuccessful. If it is successful, then the humans could land with a greater safety margin on a prepared landing surface on the second flight opportunity. Either way the site preparation capability could be included in the design of an ISRU excavator that is also designed to extract ice from the martian regolith. For example, the excavator may level a landing site and then compact it through the removal of loose surface material followed by tamping. The excavator may also be equipped with a microwave transmitter to glassify the surface of the soil to a desired depth of penetration, creating the landing pad. Landing site preparation of this sort would require the capability for precision landing to make use of the prepared site.

Presently, we could not efficiently build a landing pad on Mars and so we need to develop this technology during the lunar program. The 16-to-40 minute round-trip time-delay at the speed of light makes joystick control of the excavator impossible. We understand from conversations with experts in the mining and excavation communities that autonomous excavation is not presently possible. Efforts have been made to develop that capability, but so far they have not been successful. Simple tasks such as deciding how to push a pile of dirt will need to be re-thought and autonomous techniques will need to be developed. This will be further complicated by the pervasive dust that can get into mechanisms and cause friction, the harsh thermal environment, the lack of opportunity for maintenance and troubleshooting, and the need for long-term reliability despite these other factors. These, too, will force a re-thinking of how to do simple excavation tasks, and we may need to iterate the hardware designs through a few generations before we can get them working reliably. Finally, the lower gravity will be challenging because it will be difficult to get large reaction forces off the surface when pushing large quantities of soil, and although gravity is reduced neither momentum nor angular momentum is, so that there will be greater risk of hardware tipping over or becoming stuck.

The Moon is an excellent location to overcome these challenges. The Moon (like Mars) is dusty, has harsh thermal conditions, and has reduced gravity. In fact, the Moon is a worse case than Mars in all three regards, so excavation hardware developed on the Moon should have the greater reliability we will need when working at the vast distances to Mars. The most important thing is that the Moon is close and astronauts will be there so that when things do not work they can troubleshoot and find out why, fix things, and try again without waiting two years for the next flight opportunity. There is no facility on Earth that can provide an environment like the Moon with a vast regolith bed under vacuum conditions, not to mention the low gravity, and there is no place else in space that provides such a convenient location for martian technology development.

Sintering or otherwise stabilizing the soil will also be a challenge. Not only must we develop methods to do the actual sintering, but we must develop techniques to autonomously move the sintering unit across a large area in the harsh environment while it operates over a long period of time. There is no need for any such capability on Earth, so this requires brand-new technology and there will be a host of unexpected problems that need to be worked out. When astronauts fly to Mars, their lives will depend upon the integrity of the sintered surface and it must be certified prior to flight. The analogous processes in certifying terrestrial launch pads is much easier because mankind has a long history of working with the materials available on Earth (concrete) in the well-known terrestrial environment. The best pathway to certifying a martian landing pad will be to make similar landing pads on the Moon where they can be examined and tested in detail by the astronauts who will be there, and to dissect portions of the landing pads and return them to Earth for further analysis.

Finally, a launching or landing pad may also need navigation beacons, lighting for crew visibility, and cameras to document any unexpected conditions that may be caused by the blast of the plume. When a piece of a spacecraft is blown loose or a piece of ejecta is liberated from the landing site, it must be tracked to ensure that it does not strike critical hardware, or if it does to identify the damage as quickly as possible. Having cameras to see the problem will enable both corrective action and recurrence control. Realistic requirements for martian landing infrastructure can be hammered out at a lunar spaceport if we treat the lunar landings not just as a way to get to the Moon, but as development tests and operational practice for martian spaceport operations. This will require a mind-set that we will do more than the minimum effort to successfully land on the Moon. Lunar missions must not be entirely an end in themselves.

## VII. Conclusions

The calculations in this paper are all order-of-magnitude estimates because the numerical tools required for better predictions are still being developed. If a landing pad is not prepared on Mars, then a minimum blast radius of 1 km should be sufficient to protect hardware around the landing vehicle. However, this radius should be increased by some safety factor due to the inherent uncertainties. The blast radius may be reduced if the hardware is shielded to protect against ejecta impacts. The ejecta will have increasingly higher momentum and energy at closer distances. Also, without a prepared landing site, the bottom of the lander will be subjected to large ejecta strikes due to the presence of a narrow crater that will form under each engine, redirecting the exhaust jets back up toward the lander. Therefore, the lander will need to have its nozzles and other features adequately shielded or ruggedized to prevent loss of mission and loss of life. Furthermore, the ground under the lander will present some localized instability at engine cutoff as the excavated hole under each engine collapses and the fluidized soil returns to a solid state. The extent of this instability for cohesionless soil is predicted to be, at a minimum, the same diameter as a nominal lander. Thus, specific design work to assure lander stability after engine cutoff will be required in the future.

The alternative, which may provide greater positive safety margin as well as maintain greater payload mass to the surface, is to robotically clear and stabilize the landing site prior to the arrival of the lander. We believe this is the only alternative that is truly safe. Without an *in situ* landing pad the scale of the regolith's response to the plume will be significantly larger than what we experienced in the Apollo and Viking programs. Because the regolith density and rock placement varies randomly, there will always be some uncertainty about its actual response to a plume. We cannot tell exactly how deep or wide a plume-induced crater will be, whether the chaotic motions in the crater will randomly eject a rock at larger-than-usual velocity, whether one side of the crater will cave in more extensively than the other side, or whether the plume will expose boulders or uneven bedrock that will leave the lander unleveled. Therefore, we will not have clear safety margins when landing on unprepared surfaces. Even if the scale of plume interactions is only modest and just approaches the limits where it may affect the spacecraft, we are still driven to this conclusion; it will be difficult for program managers to accept the uncertainties of plume-induced damage considering the distance to Mars and our inability to help the crew should something go wrong. On the other hand, if we are already planning to have ISRU in the architecture of Mars missions, then it may not be too difficult to put the construction of a landing pad in the architecture as well and thereby entirely eliminate the risks of plume-induced damage. In our estimation, when we assesses this with the conservatism that usually prevails in human spaceflight, the construction of *in situ* landing pads is the best option. The time to begin developing this capability is now so that the necessary technologies can be tested and perfected during the lunar missions.